    \documentstyle[proceedings,numreferences]{crckapb} 

\begin{opening}
\title{UV EMISSION AND DUST PROPERTIES OF HIGH REDSHIFT GALAXIES}

\author{D. CALZETTI}
\institute{Space Telescope Science Institute\\
           3700 San Martin Dr., Baltimore, MD 21218, U.S.A.}

\end{opening}

\runningtitle{UV EMISSION AND DUST OF HIGH-Z GALAXIES}

\begin{document}

\begin{abstract}
The high-redshift (z$>$2) galaxies discovered over the last few years
with the Lyman-break technique represent, in number density, a major
fraction of the galaxies known in the Local Universe. Thus,
understanding the properties and the nature of these high-redshift
systems is instrumental to our understanding of the cosmic evolution
of galaxies and their stellar content. I briefly review the observed
characteristics of the Lyman-break galaxies, relate these galaxies to
their most likely low-redshift counterparts, and discuss the
implications of dust obscuration on the global properties of the
Lyman-break population.  Finally, the observational properties of the
high-redshift population are set in the framework of a simple
evolutionary model for the stellar, metal and dust content of
galaxies, to derive the {\em intrinsic} star formation history of the
Universe.
\end{abstract}

\section{Introduction}

Althought this Meeting is devoted to the discussion of the nature of
the Ultraluminous Far-IR Galaxies at low and high redshift, and their
role in galaxy evolution, I will not directly tackle this topic in my
presentation. I will instead summarize and discuss the properties of a
complementary sample of galaxies: the high-z (z$>$2), {\em UV-bright}
systems (\cite{steideletal96}, \cite{steideletal98}). The number of
known z$>$2 galaxies is now large enough that they can be classified
as a {\em population}, and have been used to infer the past star
formation history of the Universe (\cite{madauetal96},
\cite{madauetal98}).  As with all statistical studies, observational
incomplenetesses and selection biases are a concern. In the case of
the high-z population, volume corrections, luminosity selections and,
last but potentially the most important, dust obscuration effects have
been discussed by a number of authors (e.g., \cite{meureretal97},
\cite{calzetti97a}, \cite{pettinietal98}, \cite{steideletal98},
\cite{calzettietal98}). Here I will highlight the impact of dust
obscuration on the high-z galaxies and the inferred star formation
history. I hope in this way to set a ground for comparison of the
high-z UV-bright galaxy population with the high-z Ultraluminous
Far-IR galaxies recently discovered with SCUBA (\cite{smailetal97},
\cite{hughesetal98}, \cite{bargeretal98}, \cite{ealesetal98}; see also
the contributions of I. Smail, of A. Blain, and of M. Rowan-Robinson to
these Proceedings).

\section{The Lyman-Break Galaxies and their Low-Redshift Counterparts}

Since it first was presented (\cite{steideletal93}), the Lyman-break
technique had stood out as one of the most powerful tools for
identifing high-z galaxy candidates. As of this writing, more than 550
candidates have been spectroscopically confirmed to be at z$\sim$3
over an area $\sim$0.3~square degrees and about 50 at z$\sim$4 over an
area $\sim$0.23~square degrees (\cite{steideletal98}). In number
density, the bright ends of the z$\sim$3 and z$\sim$4 populations have  
similar values to the local L$>$L$^*$ galaxies, for a flat cosmology. Even
if merging may have played a role in changing these values over time,
just the number of stars contained in Lyman-break galaxies at z$>$2
accounts for $\sim$20--30\%~ of all the stars known today. The basic
fact is that the Lyman-break galaxy populations are a significant
fraction of the total galaxy population today. Thus, understanding the
nature of the Lyman-break galaxies remains a gateway to understanding
the evolution of galaxies.

The identification of high-z candidates is based on the detection of
the Lyman break at 912~\AA, which is the strongest discontinuity in
the stellar continuum of star-forming galaxies. A galaxy at, say, z=3
will have the Lyman break redshifted to 3648~\AA. If a pair of filters
is chosen to straddle the break, the galaxy will appear extremely red
in this color.  In order to avoid as much as possible low-z
interlopers, one or more filters are generally added longward of the
Lyman break to select only candidates which are blue in this(these)
additional color(s). With a careful selection of the color criteria,
the Lyman-break technique is extremely successful at identifying
high-z candidates; spectroscopic confirmations give a $\sim$95\%~ success
rate for the z$\sim$3 sample and a $\sim$80\%~ success rate for the
z$\sim$4 sample (\cite{steideletal96}, \cite{dickinson98},
\cite{steideletal98}). The lower success rate at z$\sim$4 is due to
the incidence of low-z interlopers, namely elliptical galaxies at
z$\sim$0.5--1 whose 4,000~\AA~ break falls inside the selection window
of the filters.

While the determination of the intrinsic nature of the Lyman-break
galaxies, whether they are massive systems or galaxy fragments, and
what kind of progenitors they are, is still a source of heated debate
(e.g., \cite{lowenthaletal97}, \cite{giavaletal98},
\cite{adelberger98}), the identification of their {\em observational}
low-z counterparts appears less controversial.

By selection, Lyman-break galaxies are UV-bright, actively
star-forming systems, with a preferentially blue spectral energy
distribution (SED). Observed star formation rates (SFRs) range from a
few to 50~M$_{\odot}$~yr$^{-1}$, for a Salpeter Initial Mass Function
(IMF) in the range 0.1--100~M$_{\odot}$ (\cite{dickinson98}). This
range of values is typical of what observed in Local, UV-bright
starburst galaxies (e.g., \cite{calzetti97b}). The restframe UV and
B-band half-light radii are around 0.2--0.3~arcsec, which correspond
to spatial radii $\sim$1--3~h$_{50}^{-1}$~kpc, depending on q$_o$
(\cite{giavaletal96}, \cite{giavaletal99}). The similarity of the
half-light radii at both UV and B suggests that the UV is a reliable
tracer of the full extent of the light-emitting body. Ground-based
optical spectra, which correspond to the restframe 900-1800~\AA~ range
for a z$\sim$3 galaxy, show a wealth of absorption features, and
sometime P-Cygni profiles in the CIV(1550~\AA) line (cf. the figures
in \cite{steideletal96}), typical of the predominance of young,
massive stars in the UV spectrum. Currently limited ground-based
near-IR spectroscopy (e.g. \cite{pettinietal98}) has revealed nebular
line emission in these galaxies. Hybrid line equivalent widths
constructed using the UV flux density f(UV) as denominator, namely,
EW'(\AA)=F(line)/f(UV), show that the observed values for the high-z
galaxies fall in the loci observed for local starburst galaxies
(\cite{meureretal98}). In summary, {\em the observational properties
of the Lyman-break galaxy population fully resemble, in the restframe
UV-optical range, those of low-redshift, UV-bright starburst
galaxies} (\cite{meureretal98}).

The Lyman-break galaxies share another global characteristic with the
Local starbursts. If we parametrize the observed UV stellar continuum
with a power law, F($\lambda$)$\propto\lambda^{\beta}$, Lyman-break
galaxies cover a large range of $\beta$ values, roughly from $-$3 to
0.4, namely from very blue to moderately red (Figure~1, left
panel). This range is not very different from that covered by the
Local, UV-bright starbursts (Figure~1, right panel). Population 
synthesis models (e.g. \cite{leithereretal95}) indicate that a 
dust-free, young starburst or constant star-formation population have 
invariably values of $\beta<-$2.0, for a vast range of metallicities. 
{\em What does cause the UV stellar continuum of Lyman-break galaxies 
to be redder than expected for a young star-forming population?}

\section{Dust Reddening and Obscuration in Local Starbursts and in 
Lyman-break Galaxies}

There are two main causes for a red UV stellar continuum: 1) ageing of
the stellar population; 2) presence of dust (variations of the
intrisic IMF will not be discussed here).

An ageing stellar population loses the high-mass, hot stars first, and
then, progressively, lower-mass and colder stars. In the process, the
UV stellar continuum becomes redder and redder and, also, the
4,000~\AA~ break increases in strenght. This break spans a small
wavelength range, thus is unaffected by dust reddening. The strenght
of the 4,000~\AA~ break is therefore a powerful constraint on the age
of the stellar population. The Local starbursts can have very red UV
continua ($\beta>$0), while still showing rather small 4,000~\AA~
breaks, telltales of the presence of a young stellar population
($<$10$^7$~yr) or of constant star formation (over timescales
$\sim$10$^9$~yr, see \cite{calzetti97b}). Ageing of the stellar
population is not the main reason for the presence of a red UV SED in
Local starbursts. Broad-band J, H, and K observations provide limited
information on the strenght of the 4,000~\AA~ break in high-z
galaxies, still accurate enough to exclude ageing as a general cause
for the red UV spectra in this case as well (Dickinson 1997,
priv. communication).

\def\putplot#1#2#3#4#5#6#7{\begin{centering} \leavevmode
\vbox to#2{\rule{0pt}{#2}}
\includegraphics{#1}
\end{centering}}

\begin{figure}
\putplot{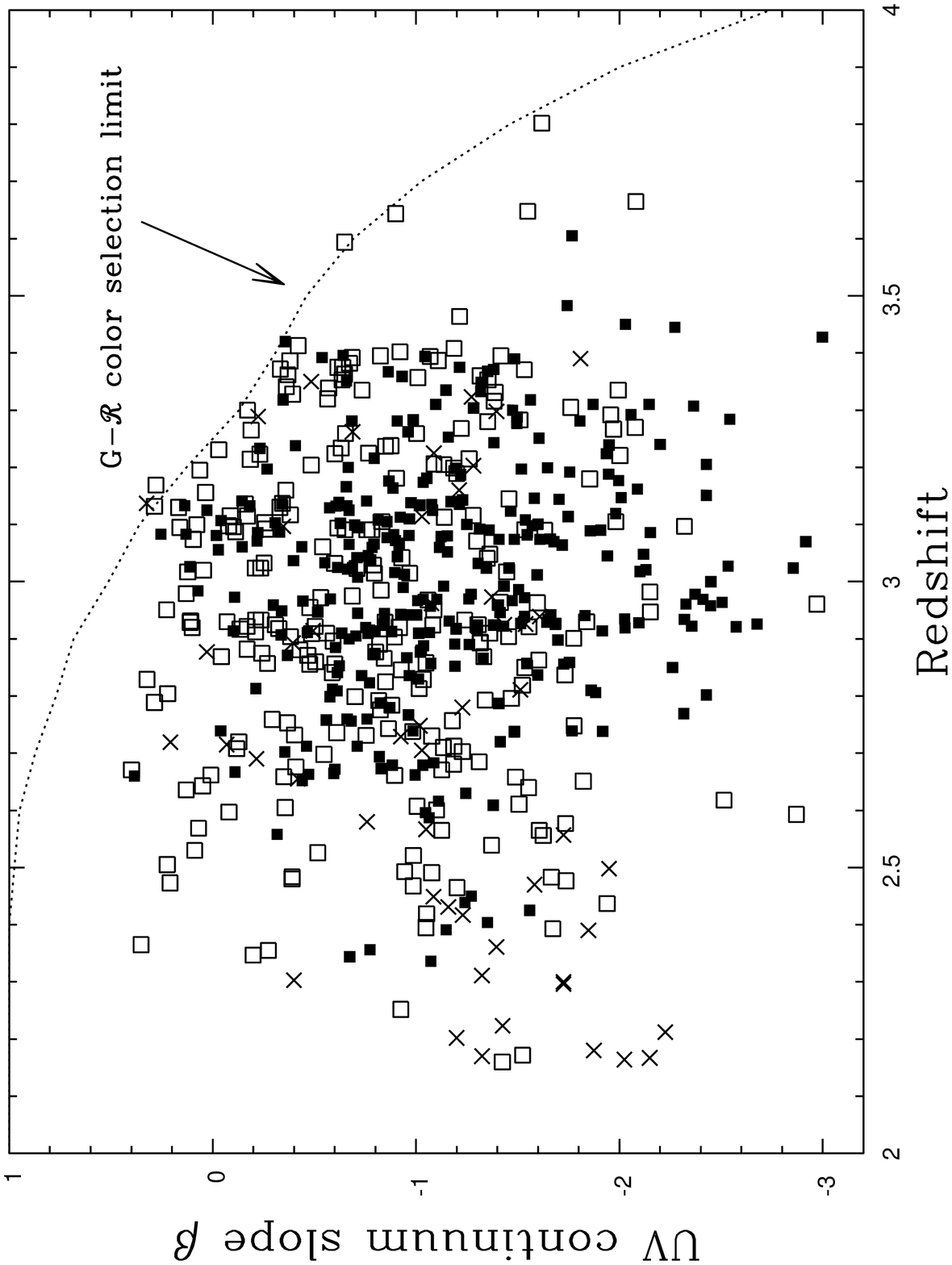}{2.5 cm}{-90}{29}{29}{-30}{105}
\putplot{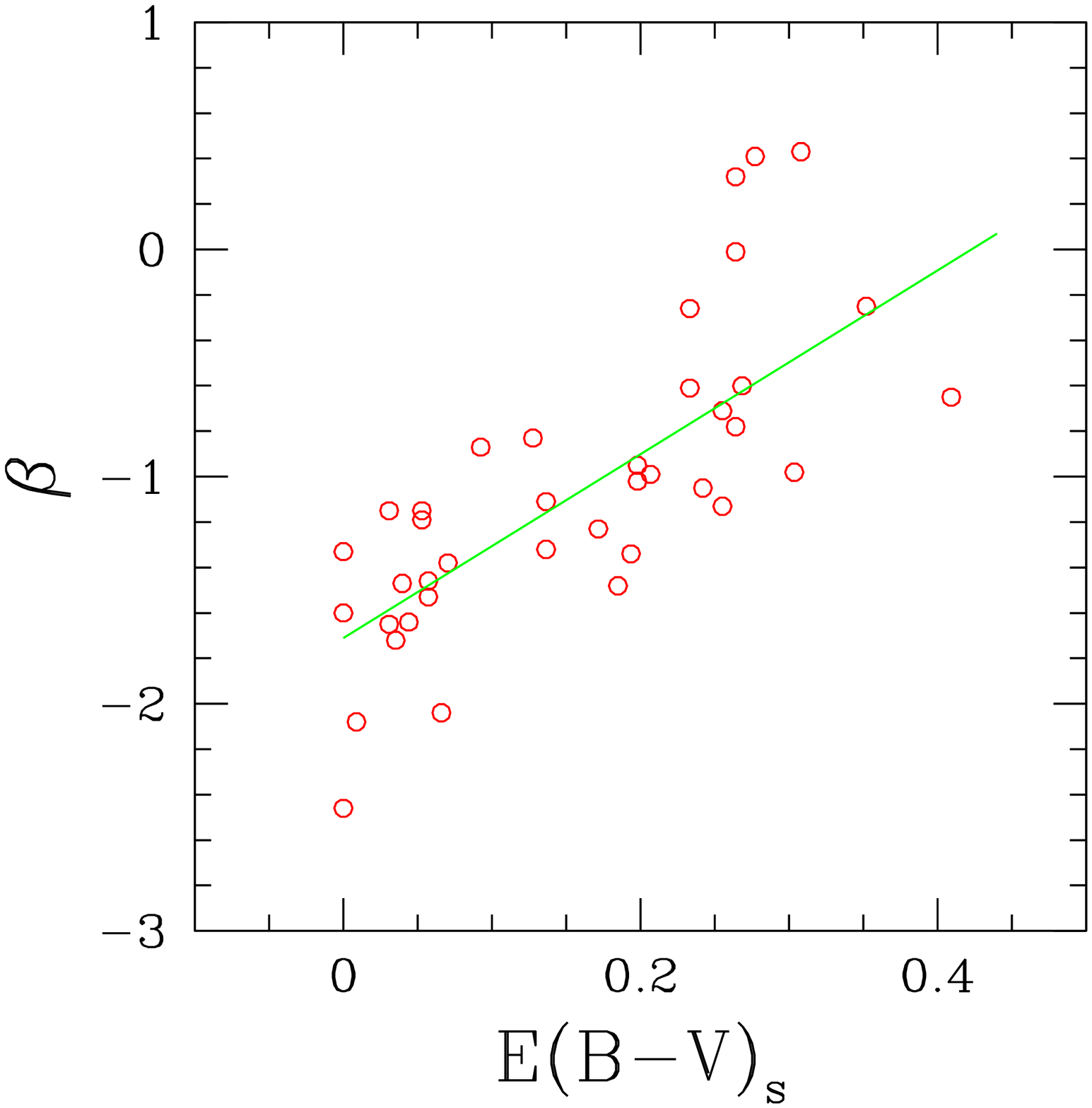}{2.5 cm}{0}{30}{30}{190}{-33} 
\scriptsize
{\bf Figure 1.} 
(Left Panel) The distribution of the observed UV
spectral indices $\beta$ of the z$\simeq$3 galaxies (M. Dickinson
1998, private communication). Only a small fraction of all the galaxies
are blue enough ($\beta<-$2) to be classifiable as dust-free, young 
star-forming populations. Different symbols corrsponds to different
color selections. (Right Panel) The distribution of $\beta$ for a
UV-bright sample of Local starburst galaxies
(\cite{kinneyetal93}). Note that the range of $\beta$ values is
similar to the high-z sample. The UV spectral slopes of the local
starbursts correlate with E$_s$(B$-$V), the color excess of the
optical stellar continuum due to dust reddening
(\cite{calzettietal94}, \cite{calzetti97b}). Blue UV spectra
correspond to small values of the color excess, red UV spectra to
large values of E$_s$(B$-$V). The best fit line through the
data is shown.
\normalsize
\end{figure}

Dust reddening is then the likely cause for red UV spectra, as
demonstrated by the correlation between $\beta$ and color excess
(Figure~1, right panel). Dust reddening is generally a
close-to-unsolvable problem for unresolved stellar populations (e.g.,
distant galaxies), because the effective obscuration will be a
combination of dust distribution relative to the emitters, scattering,
and environment-dependence of the extinction (\cite{wittetal92},
\cite{calzettietal94}). The situation gets better in the case of
starbursts because the high energy environment is generally
inhospitable to dust. Shocks from supernovae can destroy dust grains,
while gas outflows can eject significant amounts of interstellar gas
and dust from the site of star formation. If little diffuse dust is
present within the star-forming region, the main source of opacity
will come from the dust surrounding the region. Parametrizing the
`net' obscuration of the stellar continuum as: $F_{obs}(\lambda) =
F_0(\lambda)\ 10^{-0.4 E_s(B-V)\ k(\lambda)}$, with
F$_{obs}$($\lambda$) and F$_0$($\lambda$) the observed and intrinsic
fluxes, respectively, obscuration in starbursts is expressed as: 
\begin{eqnarray}
k(\lambda) &=& 2.656\, (-2.310 + 1.315/\lambda)+4.88 \ \ \ \ \ \ \ 0.63\ \mu m \le \lambda \le 1.60\ \mu m \nonumber \\
           &=& 2.656\, (-2.156 + 1.509/\lambda - 0.198/\lambda^2 + 0.011/\lambda^3) + 4.88 \nonumber \\
           & &\ \ \ \ \ \ \ \ \ \ \ \ \ \ \ \ \ \ \ \ \ \ \ \ \ \ \ \ \ \ \ \ \ \ \ \ \ \ \ \ \ \ \ 0.12\ \mu m \le \lambda < 0.63\ \mu m.
\end{eqnarray}
The connection between the color excess E$_s$(B$-$V) and the 
measured spectral slope $\beta$ is given by the correlation in 
Figure~1 (right panel). 

It is worth stressing that, although dust reddening corrections for
starbursts are parametrized above as a foreground dust screen, {\em
Equation~1 has been derived with NO assumptions on the geometrical
distribution of the dust within the galaxies}. Equation~1 is a purely
empirical result (\cite{calzettietal94}, \cite{calzetti97b}), which
includes into a single expression any effect of dust geometry, scattering,
and environment-dependence of the dust composition.

Equation~1 provides a recipe for correcting the observed SEDs for the
effects of dust reddening. Does it fully account for the dust
obscuration as well?  In other words, does Equation~1 completely
recover the light from the region of star formation or does it misses
the flux from dust-enshrouded regions? The answer to these questions
is a positive one: Equation~1 is able to recover, within a factor
$\sim$2, the UV-optical light from the entire star-forming region of 
{\em UV-bright}, i.e. moderately obscured, starbursts.

We can prove the above statement by studying the FIR emission of the
local starbursts.  Dust emits in the Far-IR the stellar energy
absorbed in the UV-optical.  However, the Far-IR emission is not, by
itself, an unambiguous measure of the opacity of the galaxy, as the
intensity of the dust emission is also a function of the SFR in the
galaxy. A good measure of the total opacity of the galaxy is instead
provided by the ratio FIR/F(UV) (\cite{meureretal98}). The Far-IR flux,
FIR, and the UV flux, F(UV), are both proportional to the SFR, but
their sensitivity to dust has opposite trends: roughly, FIR increases
while F(UV) decreases for increasing amounts of dust, although the
details of the trends are dictated by the geometrical distribution of
the dust. In the assumption that the foreground dust screen 
parametrization is valid, the FIR/F(UV) ratio is related to the 
UV attenuation in magnitudes, A(UV), via (\cite{meureretal98}):
\begin{equation}
FIR/F(UV) = 1.19 \Bigl[ 10^{0.4 A(UV)} - 1 \Bigr], 
\end{equation}
where the constant value 1.19 is the combination of the ratio of the
bolometric stellar luminosity to the UV luminosity and the ratio of
the bolometric dust emission to the FIR emission. Since F(UV) and FIR
(e.g., from IRAS) are measurable in galaxies, as is $\beta$, the UV
attenuation can be related to the UV spectral slope via Equation~2
(\cite{meureretal98}). Figure~2 shows A(UV) measured at 1,600~\AA~ as
a function of $\beta$ for a sample of Local starbursts.  Overplot on
the data is the trend predicted by Equation~1, with $\beta$ related to
E$_s$(B$-$V) using Figure~1 and the limiting case $\beta_0=-$2.1 for
E$_s$(B$-$V)=0 (\cite{calzetti97b}). Equation~1 and Figure~1 have no
adjustable parameters. The agreement between the data and the
predicted trend is therefore impressive, especially if we take into
account that the latter is a recipe for {\em reddening}, and could in
principle not account for the entire dust obscuration. Discrepancies
at the low end of the locus of data points in Figure~2 are
understandable in terms of sample incompletenesses.

\begin{figure}
\vspace{1 cm}
\putplot{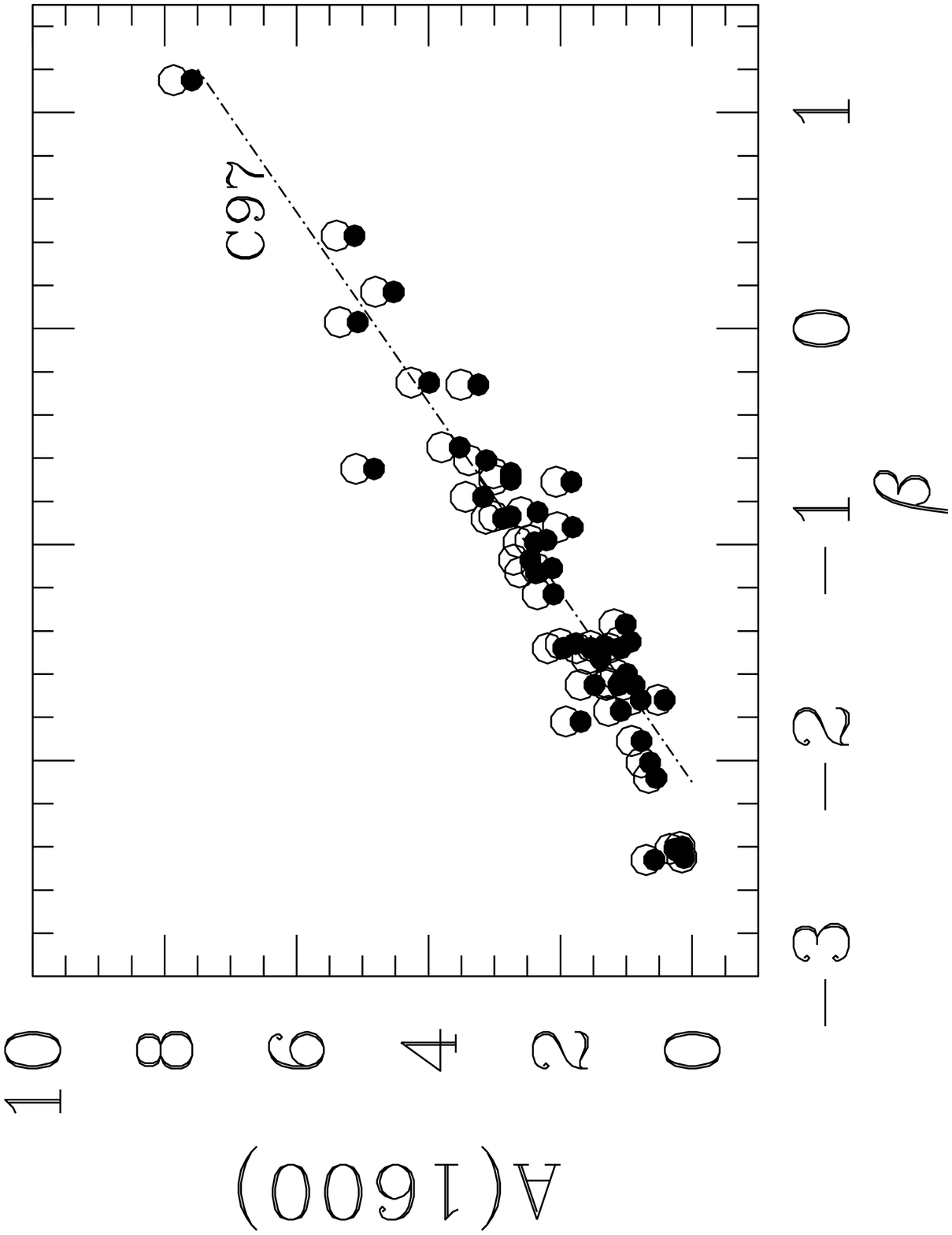}{4.5 cm}{-90}{30}{30}{55}{190} 
\scriptsize
{\bf Figure 2.} 
The attenuation in magnitudes at 1,600~\AA, A(1600), 
as a function of the UV spectral index $\beta$. The data points (empty
and filled circles) are from the same sample of UV-bright, local
starbursts of Figure~1 (\cite{kinneyetal93}). A(1,600) is proportional
to $\log [FIR/F(UV)]$ where FIR and F(UV) are the Far-IR and UV
observed luminosities, respectively (see text). The filled and empty circles
correspond to two different choices of the bolometric-to-FIR
correction for the dust emission. This is the most uncertain parameter
in the derivation of Equation~2; yet, its effect on the data is fairly
small. The straight line across the data, marked C97, is the location 
of the reddening curve of Equation~1, with $\beta_0=-2.1$ for 
E$_s$(B$-$V)=0 (\cite{calzetti97b}). There are no adjustable parameters in 
the positioning of the C97 line. The agreement between the data 
points and the predicted attenuations from the reddening curve (Equation~1) 
demonstrates that reddening corrections can fully recover the intrinsic 
UV emission from these galaxies and the amount of star formation buried 
in dust is relatively small.
\normalsize
\end{figure}

How does all this apply to Lyman-break galaxies? The entire purpose of
obtaining dust obscuration corrections for the high-z galaxy sample is
to recover the intrinsic UV emission of the galaxies, therefore
deriving a more meaningful UV luminosity function, a more accurate
value of the SFR for each object (which can bear into the
understanding of the nature of these objects), and, finally, the {\em
intrinsic} cosmic SFR density (\cite{madauetal96}). Figure~1 (right
panel) shows the observed UV spectral slopes of the z$\sim$3
galaxies. Those slopes can be `translated' into a value of the
effective color excess, which is calculated to have mean value
E$_s$(B$-$V)$\simeq$0.15 for the z$\sim$3 galaxies, or an attenuation
A(1600)$\simeq$1.6~mag (\cite{steideletal98}, see also
\cite{calzetti97a}). Incidentally, this mean value of E$_s$(B$-$V) is
similar to that observed in the local starburst sample
(\cite{calzetti97a}); this is purely coincidental, and is borne of the
fact that the two samples of galaxies cover similar ranges of
$\beta$. A similar mean value of the effective color excess has been
obtained by Pettini et al. (\cite{pettinietal98}) from the analysis of
the nebular emission lines in the NIR spectra of a small sample of
Lyman-break galaxies. Correcting the observed UV spectra for dust
attenuation increases the median SFR of $\approx\times$5 in the z$>$2
galaxies and of $\approx\times$3 in the z$\le$1 galaxies 
(Figure~3). The difference in the correction factors at low and high-z
is entirely due to the different wavelengths at which the two redshift
regimes are probed: $\sim$1,600~\AA~ the high-z galaxies and
$\sim$2,800~\AA~ the lower-z galaxies. The dust correction factors
have been `measured' only for the high-z sample and have been assumed
to hold unchanged for the z$\le$1 sample, modulo the change in
wavelength (see discussion in \cite{steideletal98}).

\begin{figure}
\vspace{1 cm}
\putplot{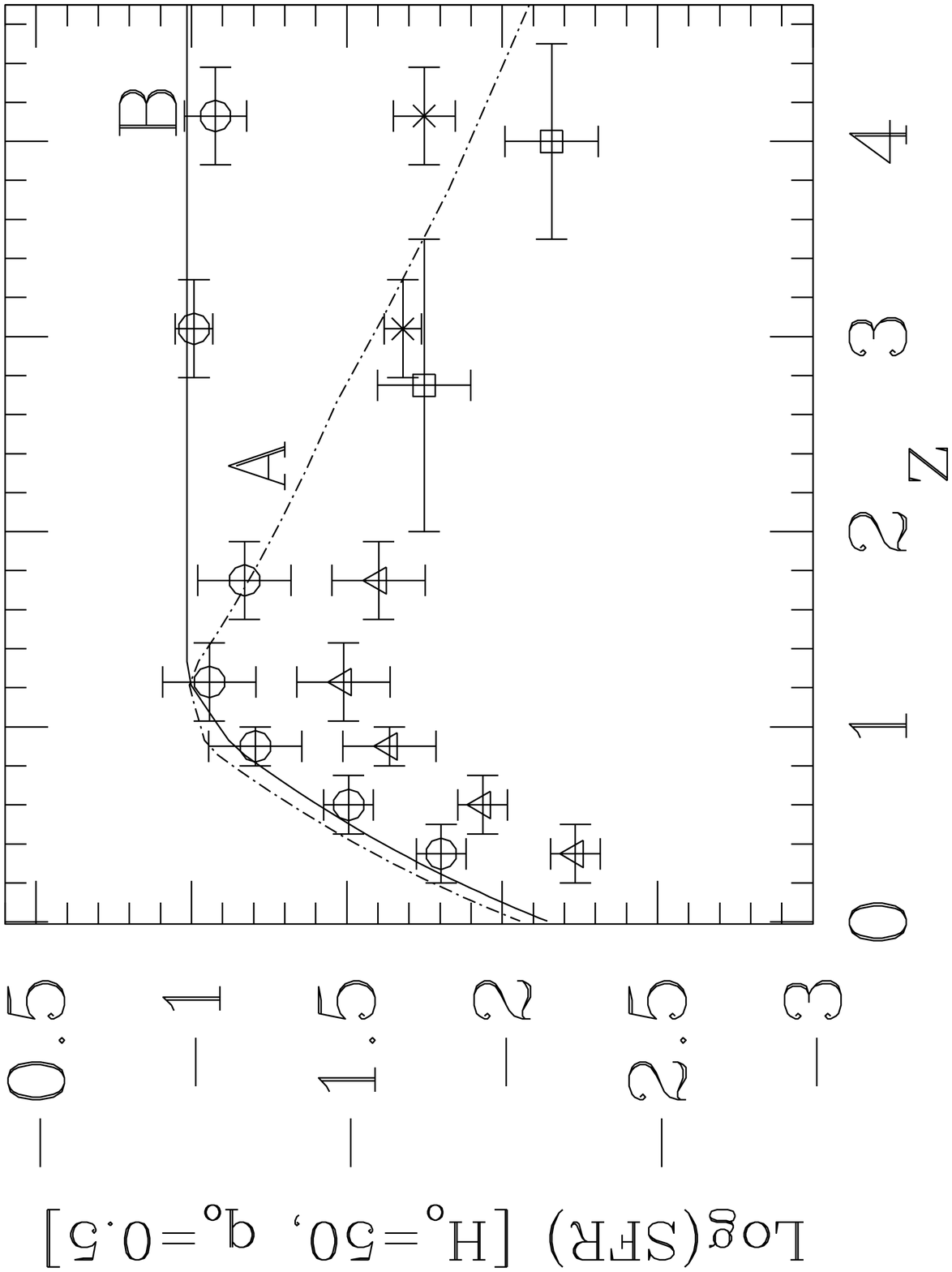}{4.5 cm}{-90}{30}{30}{55}{190} 
\scriptsize
{\bf Figure 3.} 
The SFR density, in M$_{\odot}$~yr$^{-1}$~Mpc$^{-3}$,
as a function of redshift z. The data points are derived from the UV
luminosity density of galaxies, converted to SFR using a Salpeter IMF
in the range 0.35--100~M$_{\odot}$ and continuous SF. The triangles
are the observed values at 2,800~\AA~ of Lilly et
al. (\cite{lillyetal96}) and Connolly et
al. (\cite{connollyetal97}). The squares are the observed values
at $\sim$1,600~\AA~ as reported in Madau et al. (\cite{madauetal98}),
while the crosses are the new values from the Lyman-break galaxies by
Steidel et al. (\cite{steideletal98}).  The circles represent the {\em
intrinsic} SFR densities from the data of Lilly et al., Connolly et
al., and Steidel et al., {\em corrected for dust obscuration} using
E$_s$(B$-$V)=0.15 (see \cite{steideletal98}). The curves marked A and
B bracket the range of solutions for the {\em intrinsic SFR density}
from the evolution model described in Section~4. Curve B agrees well 
with the obscuration-corrected data points (circles).
\normalsize
\end{figure}

\section{The Evolution of the Stellar, Metal, and Dust Content of Galaxies} 

The next question in line is whether a median attenuation of
$\sim$1.6~mag in the UV is reasonable at z$>2$, when galaxies where at
most a few Gyr old and, presumably, metal- and dust-poor. Both the
Cosmic Far-IR Background (CIB) detected by COBE (\cite{fixsenetal98},
\cite{hauseretal98}) and the FIR-bright galaxies detected by SCUBA at
z$\ge$1 (\cite{smailetal97}, \cite{hughesetal98}, \cite{bargeretal98},
\cite{ealesetal98}) demonstrate that dust was present at high
redshift. The luminosity of the CIB is about 2.5 times higher than the
luminosity of the UV-optical Background (\cite{madauetal98}), implying
a proportionally higher contribution of the redshift-integrated dust
emission. However, neither the CIB nor the SCUBA galaxies are telling
us how the dust content in galaxies has evolved with redshift. In the
case of the SCUBA galaxies, the redshift and luminosity distribution
and the AGN fraction of the sources will need to be tackled before
providing such information.

The time evolution of the UV luminosity density of galaxies and of the
derivative SFR density (Figure~3) can be used to constrain the metal
and dust enrichment of galaxies and, therefore, the intrinsic SFR
density (\cite{calzettietal98}). The stars which produce the observed
UV luminosity at each redshift produce also metals and dust with
negligible delay times, at most 100-200~Myr in the case of dust
(\cite{dwek98}). The obscuration from dust will produce an {\em
observed} UV flux lower than the true flux. Once the effects of the
dust on the observed UV emission are evaluated and removed, a new SFR
density is calculated. The procedure is repeated iteratively till
convergence (\cite{calzettietal98}). A number of observational
contraints are used in the model: no more than $\sim$10\%~ of the
baryons are in galaxies; inflows/outflows keep the z=0 metallicity of
the gas in galaxies to about solar, with a $\sim$15\%~ mean residual
gas content, and the z$\sim$2--3 metallicity to about 1/10--1/15 solar
(\cite{pettinietal97}); the intrinsic SFR density at z=0 must be
comparable with that measured from H$\alpha$ surveys
(\cite{gronwall98}); the dust emission must reproduce the observed CIB
and not exceed the FIR emission of local galaxies.

These constraints are still not enough to yield a unique solution; one
of the missing ingredients is the behavior of the SFR density at
z$>$4, where there are no data points. Different assumptions will lead
to different intrinsic SFR histories. The range of solutions is
bracketed by Models~A and B in Figure~3. Figure~4 shows, for each of
the two solutions, the evolution of the dust column density in the
average galaxy and the contribution to the CIB at selected wavelengths
as a function of redshift. The latter is however dependent on the
assumptions about the intrisic dust emission SED, which is not well 
constrained.

\begin{figure}
\putplot{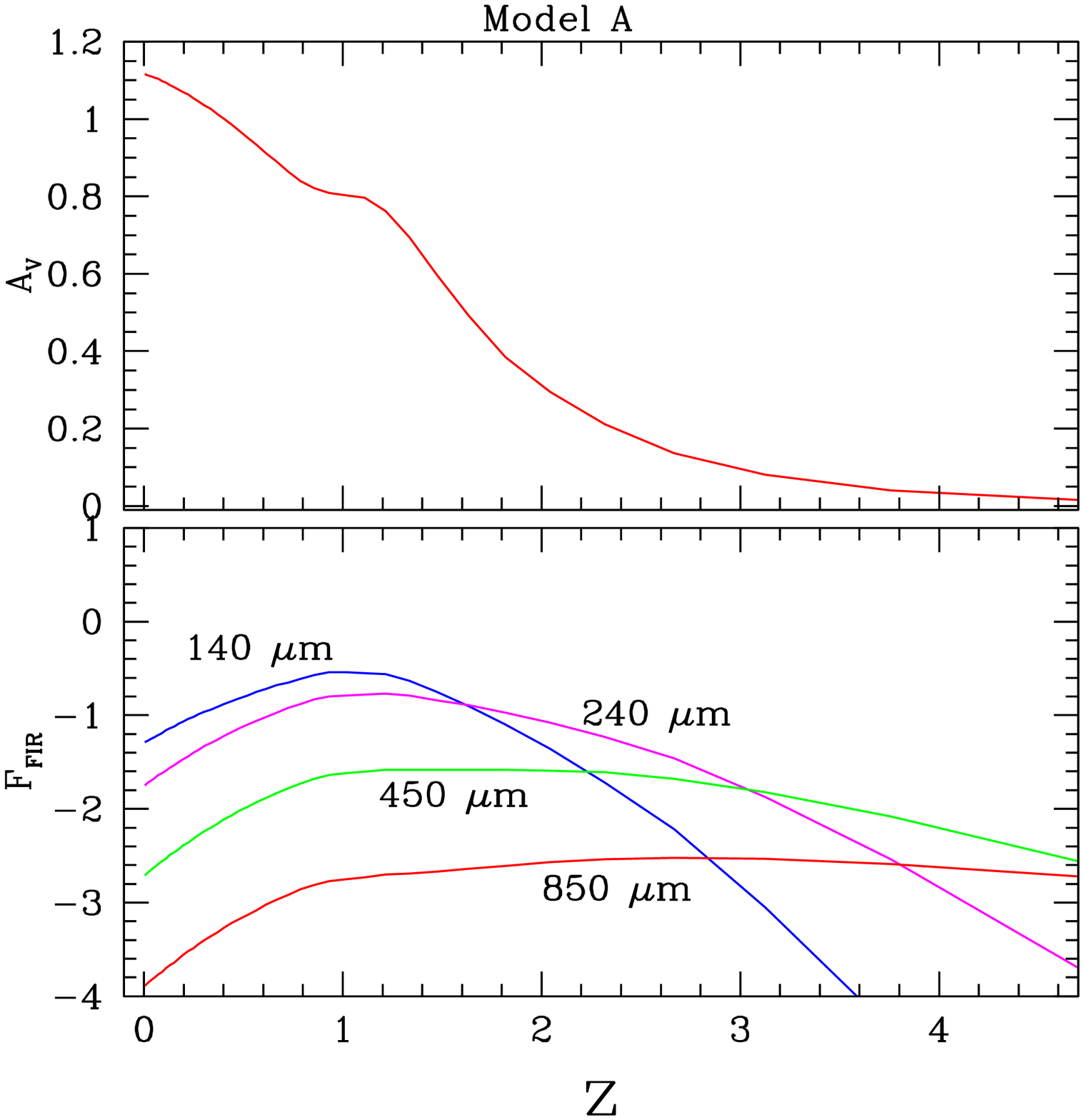}{2.5 cm}{0}{30}{30}{-10}{-115}
\putplot{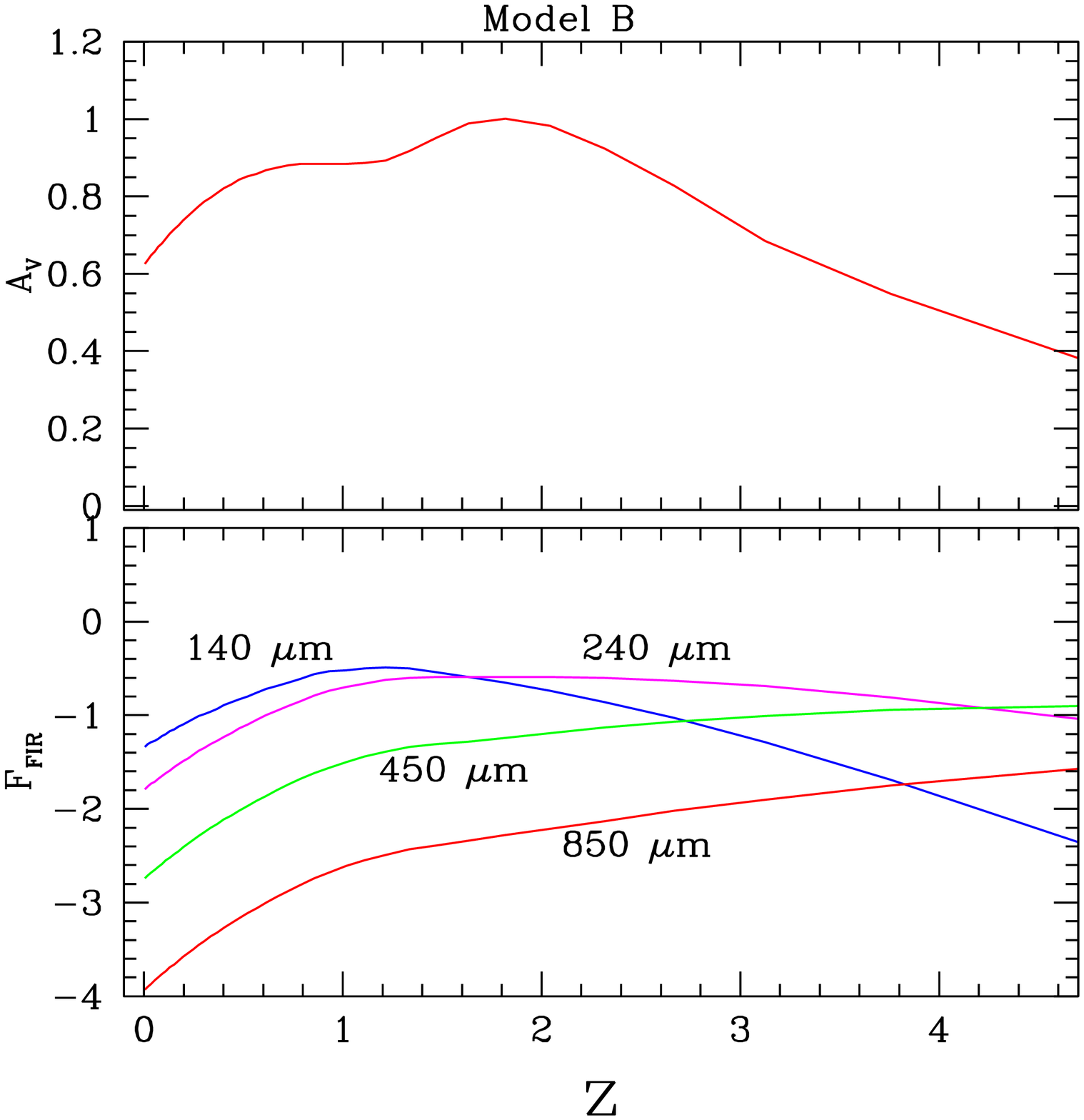}{2.5 cm}{0}{30}{30}{175}{-30} 
\scriptsize
{\bf Figure 4.} 
The evolution of the dust column density in galaxies (top panels),
expressed as optical attenuation A$_V$ in magnitudes, and of the
contribution to the CIB at selected wavelengths (bottom panels) as a
function of redshift, for both Model~A (left panels) and Model~B
(right panels). Models~A and B are the intrinsic SFR densities
described in Figure~3 and Section~4. For both models the dust optical
depth remains relatively modest at all redshifts. Yet, this is
sufficient to fully account for the CIB luminosity. The contributing
flux to the CIB is shown at the observer's restframe wavelengths
140~$\mu$m, 240~$\mu$m, 450~$\mu$m, and 850~$\mu$m, in arbitrary
units.  The two models predict different mean redshifts for the main
contributors at each wavelength. In theory, this difference could be
used to discriminate between the two solutions. However, some caution
should be used as the spectral shape of the CIB is quite dependent on
the dust emission SED adopted for the individual galaxies (in our case
a single temperature blackbody combined with a $\nu^2$ emissivity
model).
\normalsize
\end{figure}

Model~B resembles the SFR density derived from the obscuration
corrected Lyman-break galaxies (Figure~3). This demonstrates that
attenuations of about 1.6~mag in the UV are perfectly reasonable
within the framework of a simple model of stellar and dust content
evolution in galaxies.

\section{The Intrinsic Star Formation History of the Universe and the 
SCUBA galaxies}

The shape of Model~B resembles the SFR history expected for the
`monolithic collapse' model of galaxy evolution, although expectations
for hierarchical galaxy formation models are not ruled out
(\cite{whiteetal91}, see discussion in \cite{steideletal98}; see also
the contribution of G. Kauffmann to these Proceedings). Thus the
monolithic-versus-hierarchical dilemma is still unsolved by our
current knowledge of the SFR history. Values of the SFR density at
higher redshifts (z$\ge$5) will be able place more definite
constraints on the galaxy evolution scenario.

The final question we want to ask is what fraction of the total SFR
density the Lyman-break galaxies represent at each given redshift. And
how much of the SF is so deeply buried in dust that its accounting is
missing.  The obscuration curve discussed in Section~3 is technically
valid only for UV-bright star-forming galaxies; it cannot, obviously,
correct for objects which are missing from the sample because they are
too dusty. On the one hand, Model~B is only slightly in excess of the
obscuration-corrected SFR density calculated from the z$>$2 galaxies
(by a negligible amount within the observational uncertainties), and
the SFR history of Model~B is perfectly sufficient to reproduce the
observed CIB. It appears that the fraction of SF missed by considering
the Lyman-break galaxies only is relatively small. On the other hand,
a number of considerations invite to take this as a preliminary
statement.  We know that at low redshift a fraction of the star
formation is deeply buried in dust, and is obscured even at IR
wavelengths.  The same could happen at high redshift, and the SCUBA
sources seem to suggest that large dust contents are not impossible in
high-z galaxies.  The angular density of the SCUBA sources is about
1/2--1 of that of the z$\sim$3 galaxies, and are spread over a
(possibly) much larger redshift range than the Lyman-break galaxies,
namely over $\sim$5--10$\times$ larger cosmological volumes. The SCUBA
sources are then $\sim$5--20\%~ of the Lyman-break galaxies by number
density, but are forming stars with
SFR$\approx$300--500~M$_{\odot}$~yr$^{-1}$. Thus the SCUBA galaxies
could still add $\approx$25-100\% to the SFR density of the {\em
obscuration-corrected} Lyman-break galaxies, although an assessment of
the AGN contribution is still missing.

Because of their characteristics, the two populations, the UV-bright
Lyman-break galaxies and the FIR-bright SCUBA sources, are likely to
be complementary, rather than overlapping. At the level of current
knowledge, it appears that about 50--80\%~ of the SF in the early
phases of the Universe is accounted for by the {\em
obscuration-corrected} Lyman-break galaxies; the remaining 20--50\%~
of the SF may be contained in FIR-bright sources. However, more
investigation of the nature, luminosity distribution and redshift
placement of the SCUBA sources is needed before these figures can be
taken at face value.

{\bf ACKNOWLEDGEMENTS.} I am indebted with C. Steidel, M. Giavalisco,
and M. Dickinson for making their most recent results on the
Lyman-break galaxies available to me prior to publication, for
discussions, and for a critical reading of the manuscript. I would like to
thank the Organizing Committee for inviting me to this stimulating
meeting and for financially support my stay at the Ringberg Castle.

\end{document}